\documentclass[3p,times,twocolumn]{elsarticle}

\usepackage{ecrc}

\volume{00}

\firstpage{1}

\journalname{Nuclear and Particle Physics Proceedings}

\runauth{}

\jid{nppp}

\jnltitlelogo{Nuclear and Particle Physics Proceedings}

\usepackage{amssymb, amsmath}

\usepackage[figuresright]{rotating}

\begin{document}

\begin{frontmatter}



\dochead{}

\title{ABOUT MONOPOLES IN $QCD$.}


\author{Adriano Di Giacomo}
\ead{adriano.digiacomo@df.unipi.it}

\address{Dipartimento di Fisica, Universita' di PISA, 3 Largo B. Pontecorvo, 56127- Pisa, Italy. }

\begin{abstract}
The hypothesis is analysed that the monopoles condensing in $QCD$ vacuum to make it a dual superconductor are classical solutions of the equations of motion.
\end{abstract}


\end{frontmatter}


\section{Introduction.} \label{INTR}
\label{s1}
Confinement of colour is an experimental fact. No quark has ever been observed neither in ordinary matter nor as a product of high energy particle reactions.\\
An upper limit has been put by experiment to the ratio $\frac{n_q}{n_p}$ of the abundance in nature of quarks to the abundance of protons   
  \hspace{.1cm} $\frac{n_q}{n_p} \le 10^{-27}$  \hspace{.1cm}, $15$ orders of magnitude smaller than the expectation $\frac{n_q}{n_p} \approx 10^{-12}$ estimated in the Standard Cosmological Model on the hypothesis that quarks propagate as other particles \cite{okun}.\\
  Similarly the inclusive cross section $\sigma_q$ for producing a quark in a high energy reaction initiated by hadrons, which should be of the same order of magnitude as the total cross section $\sigma_T$, has an upper limit \hspace{.2cm} $\frac{\sigma_{q}}{\sigma_{T}} \le 10^{-15}$ \hspace{.01cm}.\\ The natural explanation of an inhibition by a factor  $10^{-15}$ is that confinement is an absolute property, i.e. that  $n_{q}$ and $\sigma_{q}$ are strictly zero, protected by some symmetry.  Deconfinement is then a change of symmetry.\\
  This symmetry can not be a subgroup of the gauge group, since in the Hilbert space with quarks the representation is faithful and  gauge invariance is believed to be valid both below and above the transition. It is most probably related to topological excitations on the spatial boundary at infinity, as happens in well known models of statistical mechanics in $d+1$ dimensions: in the $1+1$ dimensional Ising model
  the excitations are kinks \cite{KC}, in the $2+1$ dimensional $X-Y$ model the excitations are vortices \cite{BMK}, in the $3+1$ dimensional compact $U(1)$ gauge model on the lattice they are monopoles \cite{FM} \cite{DP}. The nature of the dual excitations is topological and is determined by $d$: they are monopoles for $d=3$.\\
  The mechanism by which $QCD$ confines color is not easy to study, since there is no precise understanding of $QCD$ at large distances.  An attractive idea is that  vacuum
  is a dual superconductor, due to condensation of magnetic charges \cite{'tH} \cite{M}: dual means that electric and magnetic are interchanged with respect to ordinary superconductors. The chromo-electric field acting between a $q-\bar q$ pair  is constrained into an Abrikosov flux tube, producing a potential proportional to the distance, in the same way as is the magnetic flux in an ordinary superconductor. Such a mechanism includes both the idea that deconfinement is a change of symmetry, and that the degrees of freedom involved are topological excitations at the spatial boundary.\\
  The mechanism has been tested in detail in $N=2 $ supersymmetric QCD \cite{SW} and in compact $U(1)$ gauge theory on the lattice\cite{FM}\cite{DP}. Unfortunately real $QCD$ is not supersymmetric, and the compact $U(1)$ gauge theory is probably not even a field theory. When it comes to physical world things become more complicated.\\
  Lattice is a non perturbative formulation of $QCD$ allowing to explore large distances. Monopoles can be detected in the lattice configurations by the recipe of Ref. \cite{'tH2}.  Having in mind  the gauge group $SU(2)$ for the sake of simplicity, the idea is to fix a gauge [Abelian Projection]  and to look at the flux through the surface of  elementary cubes of the diagonal component (3 component in colour space)
  of the magnetic field. A monopole is detected as a Dirac string, i.e. as a $2 \pi$ excess of the flux through a face of the cube \cite{dgt}.  Both the positions and the number of the observed monopoles are gauge dependent. A particular gauge,
  the Maximal Abelian Gauge \cite{'tH2}, has been privileged in the literature because the field configurations are dominated by the monopoles \cite{suz}. However an attempt to determine from the monopoles observed in lattice configurations an effective action with a minimum at a non-zero value of the monopole field was not successful\cite{pol} \cite{ts}. An alternative approach was to construct a magnetically charged operator and use its vacuum expectation value  as an order parameter for condensation \cite{dlmp} \cite{DP}.  The result was positive, showing that in  the confined phase vacuum is a dual superconductor and normal in deconfined, but the result was a kind of average over the Abelian Projections.\\
  In this work we have a different approach: we look for monopoles as classical stable field configurations and analyse their observable effects, and their possible condensation.
  \section{Monopoles in Gauge Theories.}
  Monopoles are topological configurations in $3 d$, i.e. non-trivial mappings of the sphere $S_2$ at spatial infinity onto $SO(3)/U(1)$. They are stable against local fluctuations. They can coexist if their topologies are compatible \cite{Col}.
  The prototype monopole was discovered in the $SO(3)$ Higgs model \cite{'tH1}\cite{Pol}.
  \begin{equation}
  L =-\frac{1}{4}\vec G_{\mu \nu} \vec G_{\mu \nu} +\frac{1}{2}(D_{\mu}\vec \Phi)^2 -\frac{\lambda}{2}(\vec \Phi^2- \mu^2)^2 \nonumber
  \end{equation} 
  $\mu ^2 \equiv \frac{m^2}{\lambda}$ . For a static solution $\partial_{0} =0$. In particular $\partial_{0} A_{0}=0$, which we shall assume as a choice of  gauge. The static equations of motion are
  \begin{eqnarray}
   D_{j}\vec G_{j i}&= &g \vec \Phi \wedge D_{i} \vec \Phi  - g\vec A_{0} \wedge D_{i} \vec A_{0} \nonumber \\
 D_{i} D_{i} \vec A_{0}& = &g \vec  \Phi \wedge D_{0} \vec \Phi    \label{stat} \\
D_{\mu} D_{\mu} \vec \Phi& -& \lambda \vec \Phi (\vec \Phi^2- \mu^2) =0 \nonumber
\end{eqnarray}
In presence of a non zero Higgs field a solution exists with $\vec A_{0}=0$ obeying the equations \cite{'tH1}\cite{Pol}
\begin{equation}\label{tP}
\begin{aligned}
& D_{j}\vec G_{j i} =g \vec \Phi \wedge D_{i} \vec \Phi  \\ 
& D_{i} D_{i} \vec \Phi + \lambda \vec \Phi (\vec \Phi^2- \mu^2) =0 
\end{aligned}
\end{equation}
In $QCD$ there is no Higgs field, $\vec \Phi=0$ and Eq's (\ref{stat}) read
\begin{equation}\label{qcd}
\begin{aligned}
&D_{j}\vec G_{j i} = - g \vec A _{0} \wedge D_{i} \vec A_{0}  \\
&D_{i}D_{i} \vec A_{0}=0 
\end{aligned}
\end{equation}
The minus sign in Eq(\ref{qcd}) comes from Minkowski metrics. In terms of $A_{4} = i A_{0}$ Eq's(\ref{qcd}) have the same form as
Eq's (\ref{tP}) with $\Phi$ replaced by $A_{4} $ \cite{jz} and  $\lambda=0$.
Monopoles in $QCD$ have the same form as those of Ref. \cite{'tH1} \cite{Pol}, with $A_{4}$ playing the role of the Higgs field. $\lambda =0$ implies that $\vec E = \pm \vec H$ [Self-duality or anti-delf-duality]\cite{B} \cite{PS}. $\vec E$ is the euclidean electric field, i.e. $i$ times the physical electric field. Explicitly the solutions are, in the "hedgehog" gauge :
\begin{equation} \label{spc}
\begin{aligned}
 &A^a_{i} =  \epsilon_{ain} \frac{\hat r_n}{gr}[ 1 - K(\xi)]\\
  &A_{4}^a =\frac{\mu}{g}  \hat r ^a H(\xi)
  \end{aligned}
  \end{equation}
  \begin{equation} \label{HK}
  \begin{aligned}
&K(\xi) = \frac{\xi}{\sinh\xi}\\ 
&H(\xi) = \coth\xi -\frac{1}{\xi}
\end{aligned}
\end{equation}
Here $ \xi= \mu r$; the upper index runs on colour, the lower index on space-time components. The solution is a mapping of the surface at infinity $S_2$ onto $SO(3)/U(1)$. $K(0)=1$, $H(0)=0$, $K(\infty) =0$, $H(\infty)=1$.
There is one free parameter, $\mu$, which is related to the value of $A_{4}$, the effective Higgs field, at $r= \infty$.

The magnetic field $B^a_i$ at large distance has the form
\begin{equation}
B^a_{i}\approx _{r \to \infty}\frac{\hat r^a \hat r^i}{gr^2}
\end{equation}
The direction of   $A_{4}$ in color space can be made $\vec r$ independent by a time independent gauge transformation $U(\vec r)$ transforming to the so-called Unitary Gauge.
The transformation is singular at $\vec r =0$. The result is, denoting by $T^a$ the $a-th$ gauge group generator,  $A_{4} \equiv A^a_{4} T^a$
\begin{eqnarray}
&A_{4}& = T^3 \frac{\mu}{g} H(\xi) \label{ainf} \\
&\vec  B &\approx_{r \to \infty}  T^3 [ \frac{\hat r}{gr^2} + Dirac - String]
\end{eqnarray}
The Abelian projection appropriate to expose stable monopoles in $QCD$ is the gauge $\partial_{4} A_{4} =0$ and $A_{4}$ diagonal.
This gauge is perfectly defined and has no gauge freedom left except the residual $U(1)$. \\
$A_{4}$ in this gauge  can be traded with the Polyakov line $L(\vec r)$, which is defined as the trace of parallel transport along the time 
axis with periodic boundary conditions in time, is gauge covariant and hence its trace is gauge invariant.
\begin{equation}
L(\vec r) = P \exp( \int^{\frac{1}{T}}_0 i gA_{4}(\vec r, \tau) d \tau)  \label{pl}
\end{equation}
Here $P$ denotes ordered product, $T$ is the temperature or $\frac{1}{T}$ the extension of the system in the temporal direction.
$QCD$ will be defined as the limit for $T\to 0$. 

In the gauge $\partial_{4} A_{4} =0$ Eq.(\ref{pl}) simplifies to 
\begin{equation}
L(\vec r) = \exp(i \frac{gA_4(\vec r)}{T}) \label{polline}
\end{equation}
The gauge in which $A_{4}$ is time independent and diagonal coincides with the gauge in which the Polyakov line is. The appropriate abelian projection for monopoles as stable classical solutions is the one in which the Polyakov line is diagonal. 

Notice that Eq.'s (\ref{qcd}) are invariant  under the change  $A_{i} \to A_{i}$ $A_{4} \to -A_{4}$, and, as $QCD$, under parity and charge conjugation. By invariance under  parity [ $A_{i} (\vec r) \to - A_{i} (-\vec r)$ , $ A_{0} (\vec r) \to A_{0}( -\vec  r)$] for each solution there will be a solution with the same energy having the same magnetic field $\vec B$ and opposite electric field $- \vec E$ and the same boundary condition at $r \to \infty$.\\ Under  charge conjugation, in the usual representation for $SU(2)$ with $T_{1}$, $T_{3}$ real and $T_{2}$ immaginary the charge conjugate classical solution is minus the complex conjugate, which is of course again a solution with the same energy.    
Changing $A_{4} \to - A_{4}$, which is also a symmetry, we get a solution with the original boundary conditions, the same electric field and opposite magnetic field. 
The energy (mass) of the monopole is \cite{'tH1} $ M = \frac{4 \pi}{g^2} \mu$.\\

Everything is easily generalized to a generic gauge group, but we will only consider  $SU(N)$.\\
For $SU(N)$ there exists one $SU(2)$ subgroup for each of the $\frac{N(N-1)}{2}$ positive roots of the Lie algebra, and with it a static monopole solution.
For the $(N-1)$ simple roots  $\vec \alpha_{i} $  , $i = 1...(N-1)$ the solutions read, in the unitary gauge,
\begin{equation}
A^{(i)}_4 = \mu [C - T^{(i)}_3 (1- H(\xi))]  \label{gen}
\end{equation}
Here $T^{(i)}_3$ is the third component of the $SU(2)$ sub-algebra spanned by the generators corresponding to the $i-th$ root ,  $C=\Sigma_{i} C^i$ is the sum of the fundamental weights corresponding to the $i-th$ simple roots. The space components $A^{(i)}_k$ have the same form as in Eq(\ref{spc}) with the index $a$ running in the appropriate $SU(2)$ sub-algebra corresponding to the $i-th$ simple root.  Recall that $[C^i, T^j_{3}] =0$, $[C^i, T^j_{\pm}] = \pm \delta^{ij}T^i_{\pm} = \delta ^{ij}[T^i_3,T^i_{\pm}]$. This makes $C - T^i_3$ commute with $T^i_{\pm}$, which is the condition for (\ref{gen}) to be a solution of Eq (\ref{qcd}). Invariance under Weil transformations fixes the scale $\mu$ to be the same for all simple roots.\\
For the roots $\vec \alpha _{(t,k)} \equiv  \vec \alpha_t +...\vec \alpha_{t+k-1} $, $t= 1,...N-k$ the solution with the same boundary conditions reads
\begin{equation}
A_4^{(t,k)}   = \mu [ C  - k T_{3}^{(t ,k)}(1- H( k\xi) ) ] \label{montk}
\end{equation}
In this notation $k=1$ corresponds to fundamental monopoles. 
The mass of the monopole $(t,k)$ is  k times the mass of the monopole for a simple root, namely  $M^{(t,k)} =k \frac{4 \pi}{g^2} \mu$.
The factor $k$ is dictated by the requirement that the boundary condition at large $r$ be the same as for the  fundamental monopoles. Since
$[C, T_{\pm}^{(t ,k)}] = \pm k  T_{\pm}^{(t ,k)}$, $[T_{3}^{(t ,k)}, T_{\pm}^{(t ,k)}] = \pm T_{\pm}^{(t ,k)}$ the factor $k$   is needed to make $[C  - k T_{3}^{(t ,k)}, T_{\pm}^{(t ,k)}] =0$ and have a solution.
 There is again only one free parameter, the scale $\mu$.\\
 Besides the above monopoles there exist their excitations\cite{Lee}. Since $\mu$ is arbitrary, it can be changed to
  $\mu_{n} \equiv  \mu + n 4 \pi T$ with $n$ an integer and the solution Eq(\ref{montk}) to $A^{(tk)}_{4} = C[ \mu + n 4 \pi T] - kT^{(t,k)}_{3} \mu_{n} (1 - H(k \xi_{n}) )$, with $\xi_{n} \equiv  r\mu_{n}$. A gauge transformation 
  \begin{equation}
  U^4_{r} (x_{4})  = \exp( i 4 \pi n T  C x_{4}) \label{gt}
  \end{equation} 
  periodic in $x_{4}$, transforms  $A^{(t,k)}_{4} \to  A^{(t,k)}_{4}  - C n 4 \pi \frac{T}{g}  $, or to $A^{(t,k)}_{4} = C \mu -k T^{(t,k)}_{3} \mu_{n} (1 - H(k\xi_{n}) )$.  A monopole with the same quantum numbers, the same boundary condition at large $\vec r$ and mass $ M^{(t,k)} +  (k 4 n \pi) \frac{T}{g^2}$.
The classical weight $\exp(-S) = \exp(-\frac{m}{T})$ is depressed by a factor  $\exp(\frac{-kn4 \pi}{g^2})$ with respect to original monopole.\\
Another series of monopole solutions  can be obtained by use of the excitations of the charge conjugate solutions.  Define 
 $\bar \mu$  by the equation 
 \begin{equation}
  k \mu + (N-k) \bar \mu = 2\pi T \label{mubarmu}
  \end{equation}
  The charge conjugate of the solution Eq(\ref{montk}) is again a solution, and, since $\mu$ is arbitrary,  
$k\mu$ can be changed  to $ (N-k) \bar \mu$ .  The resulting solution is  $ C \mu + (N-k)\bar \mu T^{(i,k)}_{3} (1 - H(\bar \xi (N-k)) )$ again compatible with the boundary conditions at spatial infinity. Indeed it is easily seen that  $C = \ C^{(t,k)} + kT^{(t,k)}_{3}$  , with $[ C^{(t,k)}, \vec T^{(t,k)}]=0$.  By use of Eq(\ref{mubarmu}) $C \mu + (N-k)T^{(t,k)}_{3} \bar \mu = \mu C^{ik} +2\pi T T^{(t,k)}_{3}$. The last term can be gauged away by a transformation periodic in time of the type Eq(\ref{gt}). $\bar \xi \equiv  \bar \mu r$.\\
In summary, neglecting excitations, in $SU(N)$ gauge theory, there are $\frac{N^2-1}{2}$ monopoles $(t,k)$, $t=1,...N-k$ and $k=1,..N-1$, with scale parameter $\mu k$, independent of $t$, $N-k$ for each value $k$, to which additional $k$ can be added
with scale parameter $\bar \mu k$, defined by the relation $\bar \mu k + \mu(N-k)= 2\pi T$ , for a total of $N^2 -1$. The corresponding  $A_{4}$'s are those of Eq(\ref{montk}) for the first set $t=1...N-k$ and, for the second set $t= N-k+1...N$
\begin{equation}
A_4^{(t,k)}   = \mu  C  - k\bar \mu T_{3}^{(t ,k)}(1- H( k\bar \xi) ) ] \label{montk1}
\end{equation}
The  generators are $N \times N$  matrices. The  $T_{3}^{(t ,k)}$ are diagonal with diagonal elements \\$[T_{3}^{(t ,k)}]_{n,n} = \delta _{n,t} - \delta_{n,t+k}$  for $ t \le (N-k)$, and\\
$[T_{3}^{(t ,k)}]_{n,n} = \delta _{n,t} - \delta_{n, t+k-N} $  for  $t > (N-k)$. \\ $ [T_{+}^{(t,k)}] _{l,m} = \delta _{l,t} \delta _{m, t+k} $ for $t \le (N-k)$ , \\ $[T_{+}^{(t,k)}] _{l,m} = \delta _{l,t} \delta _{m, t+k -N}$ for  $t > (N-k)$. \\ $T_{-}^{(t,k)} = [T_{+}^{(t,k)}]^{\dagger} $. \\
It is a general theorem that for a soliton the trace of the Polyakov line at large distances is constant on the sphere at infinity, independent of the direction\cite{GPY}. That value is called Holonomy, $\frac{1}{N} Tr[ L(\infty)]$ . In our case it can be explicitly computed giving
\begin{equation}
\frac{1}{N} Tr[ L(\infty)]= \frac{1}{N} Tr[\exp( i C \alpha)] =\frac{1}{N} \frac{\sin( \frac{\alpha N}{2})}{ \sin (\frac {\alpha}{2})} \label{Hol}
\end{equation}
 Here $\alpha \equiv \frac{\mu}{T}$  and 
\begin{equation}
C = diag ( \frac{N-1}{2}, \frac{N-1}{2} -1,...-\frac{N-1}{2}) \label{C}
\end{equation}
The holonomy Eq(\ref{Hol}) vanishes for $\alpha= \frac{2 \pi}{N}$  and is $=1$ for $\alpha= 0 \mod \frac{4 \pi}{N}$.
\section{ The Polyakov line and Confinement.}
For pure gauge theories (no quarks) the trace of the Polyakov line averaged over the position   $\langle L \rangle$ is the order parameter for confinement : it vanishes in the confined phase, and is non zero in the deconfined one.  $\langle L \rangle =\exp(-\frac{F_q}{T})$ where $T$ is the temperature, and $F_q$ the free energy of a static quark: $\langle L \rangle=0 $ means diverging $F_q$ i.e. confinement. \\
The symmetry involved is the invariance under the Centre of the gauge group, which leaves the ground state invariant  in the confined phase, and is spontaneously broken in the deconfined one. In presence of dynamical quarks that symmetry does not exist. Therefore that symmetry can not be the symmetry responsible for confinememt, as discussed in Sect.\ref{INTR} since quarks do exist in nature. However it can be related to it, and coincide with it in the quenched case as we shall argue below. Empirically the Polyakov line still works as an order parameter: it has a rapid variation at the transition, or a peak in the susceptibility, which obeys  scaling with the appropriate critical indexes as in the quenched case.\\
From the definition Eq.(\ref{polline}) it could seem that the Polyakov line has infrared problems when the temperature $T$ tends to zero,
the limit which defines $QCD$. In fact as $T \to 0$ the time extension $\Delta x_{4}$ of the system becomes larger and larger, much larger than the  correlation length $\frac{1}{\Lambda}$ , which is typically of the order of $1fm$, approximately equal to the inverse of the temperature
of deconfinement. Since at time differences larger than the correlation length the boundary conditions become irrelevant, one can conclude that as $\Delta x_{4}\to \infty$, $\langle L \rangle \approx  \langle L \rangle^k $, $k$ an integer, which implies either $\langle L \rangle=0$ (confinement) or $\langle L \rangle=1$, and that, below the deconfining transition, $ \Delta x_{4} =\frac{1}{T} $ in Eq.(\ref{polline}) can be replaced by $\frac{1}{\Lambda}$. If $\langle L \rangle$ has to be the holonomy of monopoles, in Eq.(\ref{Hol}) $T$ is replaced by $\Lambda$, 
 $\alpha =\frac{ \mu}{\Lambda}$, 
and $ \mu = \frac{2 \pi}{N} \Lambda$. The size of the monopoles is fixed by  the correlation length. \\
In the confined phase $\alpha= \frac{\mu}{\Lambda}$ in  Eq(\ref{mubarmu}) gives $\bar \mu = \mu$.
Monopoles appear in sets of $N$ with the same mass: the $N-1$ fundamental monopoles degenerate with the charge conjugate of the
monopole corresponding to the root $\vec \alpha^1 + \vec \alpha ^2+  ...\vec \alpha ^{N-1}$ have scale parameter $\mu$; The $N-2$ monopoles corresponding to the roots $\vec \alpha^1 + \vec \alpha^2$ ....$\vec \alpha^{N-2}+\vec \alpha^{N-1} $ have scale parameter $2 \mu$ and are degenerate with the charge conjugates of the monopoles corresponding to the roots $\vec \alpha^1+ \vec \alpha^2+..\vec \alpha^{N-2}$ and $\vec \alpha^2+ \vec \alpha^3+..\vec \alpha^{N-1}$, and so on till the monopole $\vec \alpha^1+ ..\vec \alpha ^{N-1}$
together with the charge conjugates of the fundamental monopoles $\vec \alpha^1$ ...$\vec \alpha^{N-1}$ which have scale $(N-1) \mu$.
The fundamental monopoles will dominate since the weight is proportional to  of $\exp(-\frac{mass}{\Lambda})$.\\
One can compute the Polyakov line distribution of the monopoles: the computation is tedious but trivial since the exact form of the monopole field is known. The contribution of a monopole {(t,k)} sitting at the origin depends on the distance from the centre and for $t=1,  (N-k)$ has the form
\begin{eqnarray}
 L^{(t,k)}(\vec r) - L_{\infty}&=&\frac{2}{N} \exp(-i \alpha [\frac{N-k+1}{2} - t]) \nonumber \\ &&[\cos(\frac{\alpha k}{2})) -\cos(\frac{\alpha k}{2}H( k \xi )]  \label{L}
\end{eqnarray}
For   $t =N-k +1,...N$
\begin{eqnarray}
     L^{(t ,k)}(\vec r)  - L_{\infty} &=&\frac{2}{N} \exp(-i \alpha [\frac{N+1}{2} - t])\exp i \frac{\bar \alpha k}{2} \nonumber \\ &&[\cos(\frac{\bar \alpha k}{2}) -\cos(\frac{\bar \alpha k}{2}H(k \bar \xi ))] \label {LBAR}
     \end{eqnarray}
    
     In the confined phase  $\bar \alpha = \alpha = \frac{2 \pi}{N}$ , $\bar \xi = \xi$ and the two equations  (\ref{L}) and (\ref{LBAR}) have the same form, with the index t running from $1$ to $N$, namely  
     \begin{equation}
      L^{(t,k)}(\vec r) - L_{\infty} =\exp(it \frac{2 \pi }{N}) \Phi(k,\xi) \label{LCONF}
      \end{equation}
      where
      \begin{eqnarray}
     \Phi(k,\xi) &=&\frac{2}{N} \exp(-i \alpha [\frac{N-k+1}{2}]) \nonumber \\ &&[\cos(\frac{\alpha k}{2} ) -\cos(\frac{\alpha k}{2}H( k \xi ))]
       \label{FI}
      \end{eqnarray}
      In the confined phase $L_{\infty}=0$ [Eq(\ref{Hol})] and the contribution of the monopole $t$, for whatever  value of $k$, is proportional to the $t-th$ $N$ root of $1$. If the weight is independent on $t$ the contributions of the monopoles sum up to zero. It is easily seen that the argument holds for all higher excitations of the monopoles. Adding monopoles to the confining vacuum leaves $\langle L \rangle$ locally unchanged.
      
      In the simple case of $SU(2)$ $k=1$ and Eq(\ref{LCONF}) has the form
      $L^{(t,1)} = \exp(i \pi t) \cos(\frac{\pi}{2} [\cosh(\xi) -\frac{1}{\xi}])$ . The two values  $t=1$ and $t=2$ give opposite sign: if the distribution is the same: this makes
      the average value  $\langle L \rangle =0$.
      
      \section{ The vacuum state.}
      If the ground state in the confined phase in the  classical limit is a field configuration of monopoles with compatible boundary conditions at large distances, the size of the monopoles is fixed to be $\mu = \frac{2 \pi}{N}\Lambda$. The shape of the monopoles Eq.(\ref{FI})
      and their distribution determines the distribution of the Polyakov line in complex space. We shall consider monopoles with $k=1$: the result is trivially extended to higher mass monopoles. The contribution of a single monopole configuration is, apart from a factor proportional to the $N$-th root of the identity [Eq.s (\ref{LCONF}) and (\ref{FI}) ],
      \begin{equation}
      L(\xi) = \frac{2}{N} [ \cos(\frac{\pi}{N}H(\xi))  - \cos(\frac{\pi}{N})] \label{DISTR}
      \end{equation}
      $H( \xi) = \coth \xi - \frac{1}{\xi}$ grows monotonically from $0$ at $\xi=0$ to $1$ at $\xi \to \infty$, and $ L$ decreases correspondingly from
     $ L_{max} =  \frac{2}{N} [- \cos(\frac{\pi}{N})  + 1]$ to $0$. At small values of $\xi$\\ $H(\xi) \approx \frac{\xi}{3}$ and 
     \begin{equation}
      \frac{1}{2} (\frac{\pi}{2N})^2 \xi^2 \approx L_{max} -L(\xi) 
     \end{equation}
     Since  $ \frac{dL}{d\xi}\approx -\xi(\frac{\pi}{2N})^2$, the distribution in space   $P(\xi) d \xi =\frac{1}{V} \frac{4 \pi}{( \mu)^3} \xi^2 d\xi$ can be traded with the distribution in $L$  $P(L) = P(\xi)/ |\frac{dL}{d\xi}|$ . If $N_M$ is the number of monopoles in the volume $V$ and $\rho _M$ their density each of them will contribute to the distribution near $L_{max}$  in the same way, giving
     \begin{equation}
     P(L) = K \frac{ 4 \pi\rho _M}{ 3 \Lambda^3}  (N[L_{max} - L])^{\frac{1}{2}} \label{PL}
     \end{equation}
with $K=\frac{N^7 3^4}{2^4  \pi^6	}$. The value $\alpha  = \frac{2 \pi}{N}$ has been used. For $SU(2)$ the distribution observed on the lattice is $P(L) \approx \frac{2}{\pi} ( 1 - L^2)^{\frac{1}{2}}$ and the number of monopoles in a cube of size of the order of the correlation length is of order unity. A systematic comparison of Eq(\ref{PL}) with the distributions observed on the lattice at various $N$ would be of interest.\\
 What we have done so far is to take seriously the idea that monopoles in $QCD$ exist as stable topological configurations and draw the consequences: the first consequence of this assumption is that the monopoles belong to the abelian projection in which the Polyakov line is diagonal. In addition they are BPS monopoles, i.e. self-dual or anti-self-dual configurations. This feature can in principle be verified in lattice configurations: an excess of magnetic flux trough a plaquette $(i, j)$ of an elementary cube should correspond to an excess of electric flux through the space-time plaquette $(0,k)$ with $ \vec k= \vec i \wedge \vec j $.\\
 We have argued that, below the deconfining transition down to zero temperature the effective time extension which determines the Polyakov 
 line is the correlation length of the system $\frac{1}{\Lambda}$. Requiring the holonomy to be that of confining vacuum, namely $Tr L =0$,
 fixes for $SU(N)$ to $\frac{2 \pi}{N}$ the ratio $\frac{\mu}{\Lambda}$ of the scale parameter of the monopoles to the physical scale $\Lambda$ of the system. \\
 Notice that any gauge transformation periodic in $x_{4}$ with period $\frac{1}{T}$ leaves the trace of the Polyakov line, unchanged.\\
 Requiring the compatibility of the boundary conditions for all the solutions fixes the mass spectrum of the monopoles. The spectrum is specially simple if the holonomy vanishes (Confinement): the monopoles group into $N$-plets and the contribution to the Polyakov line of their members differ by a factor $N$-th root of the identity, so that they sum up to zero if their distribution is symmetric. Adding a monopole leaves locally unchanged the Polyakov line, and in particular the confining vacuum.\\
This picture nicely fits the arguments of Ref.\cite{SY} on the universality class of the deconfining transition.\\
For $N \ge 3$ pure gauge the transition is first order, and the corresponding temperature $T_c \approx \Lambda$: below $T_c$ $\alpha= \frac{g \mu}{\Lambda} =\frac{2\pi}{N}$, $\langle L \rangle =0$ Eq(\ref{Hol}), $\bar \mu =\mu$. In the deconfined phase $\mu =0$, $A_{4}=0$ and there are no stable monopoles.\\

What is observed in lattice $QCD$ configurations \cite{Kanazawa1,Kanazawa2} is that, in the confined phase, stable monopoles do exist in the maximal abelian gauge, which wrap  the time direction of the lattice via periodic boundary conditions, the so-called "long monopoles". These monopoles dominate physics, the short ones are irrelevant.\\
 It is also known that  for static monopoles the unitary gauge and the maximal abelian gauge coincide \cite{bdd}. Our stable monopoles could then be identified with the long monopoles.\\
 Magnetic charge is conserved, so that monopoles of any kind describe closed loops in space-time, most of which close without crossing the lattice in the time direction. Some of them could be lattice artefacts, coming from the method of detection\cite{dgt}. Empirically all of them seem to be irrelevant to physics.\\
 A guess for the ground state  $| G \rangle$ could be a coherent state of stable monopoles
 \begin{equation}
 | G \rangle = \Pi _{r=1}^N\exp( \beta b_r^{\dagger}) | 0 \rangle \label{GST}
 \end{equation}
 with $b _r= b_r ( \vec p =0) $ the destruction operator at momentum $\vec p =0$ of the r-th  monopole with $k=1$[Eq.s(\ref{montk}), (\ref{montk1}) ]. To guarantee invariance
 the operator acting on $|0\rangle$ in  Eq(\ref{GST}) has to be multiplied by its space reflected and the result by its charge conjugate.\\
 A state of the form Eq(\ref{GST}) is left invariant, apart for a trivial factor, by applying the sum $\Sigma _{r} b^{\dagger}_r$, since the weight is simmetric in $r$, [ Eq(\ref{LCONF}) ]. In principle a linear combination of monopoles is not a monopole, since Eq(\ref{qcd}) is non linear: it is not trivial to define a Hilbert space for monopoles.
 However a scalar product can be defined in terms of the classical field configurations, , which makes monopoles with different charges orthogonal, by use of the conserved magnetic current. Physically screening of the magnetic field produced by dual superconductivity makes monopoles practically non interacting.\\
 Monopole condensation in this picture is Bose condensation as discussed in  \cite{dds}. The order parameter is
 \begin{equation}
 \beta =  \langle G| b_r(\vec 0)| G \rangle
 \end{equation}
The vacuum Eq(\ref{GST}) is only a plausible guess. Proving it is not easy.\\
An alternative approach starts from Calorons which are a generalisation of the usual instantons to the case of a non trivial Holonomy \cite{GPY} \cite{Lee}.
In Ref\cite{KvB} it was shown that they have monopoles [dyons] as constituents. Their weight can be computed as for ordinary instantons \cite{DG}. This allows to compute the weight of the monopoles and to circumvent the infrared problem that would emerge in computing it for a single monopole, which is charged. A generalisation of the instanton ensemble model to the case of non trivial holonomy. Even in that model the attempt to show that there is a minimum at $\langle L \rangle =0$ in the effective action for the Polyakov loop was not successful \cite{Diakonov}. 
 The assumption is that all the monopoles can be considered as constituents of calorons. Lattice data \cite{DGH}show that this may not be the case: in $SU(3)$ gauge theory with no quarks adding monopoles to field configurations changes the number if instantons , one for each monopole anti-monopole pair. Moreover it is known that instanton ensembles have difficulty in explaining confinement.\\
 Dual superconductivity of the vacuum can be the way to explain confinement in terms of symmetry, and in particular an alternative way to  look at the centre symmetry  in the pure gauge case. Stable monopoles seem to play an essential role. More work is needed to put it on safe grounds.
 
 \section{Aknowledgements}
Thanks are due to C. Bonati for useful discussions.

\end{document}